# Does only the amplitude of the state vector carry the physical information?

**I. G. Koprinkov**
*Department of Applied Physics, Technical University of Sofia, 1756 Sofia, Bulgaria*
*E-mail: igk@vmei.acad.bg*

**Abstract:** We present theoretical and experimental evidences, which show that the material phase of the state vector is causally related with the dynamic of the quantum system and becomes carrier of physical information.
©1999 Optical Society of America

The quantum information problem is closely related with the fundamental question of the physical information carrier. At first glance, rising up the question of the information carrier sounds meaningless because, according to the standard interpretation of quantum mechanics (QM), it is only the amplitude of the state vector becomes physical observable. Thus, although the Schröedinger equation

$$\hat{H}|\Psi(\vec{r},t)\rangle = i\hbar\partial_t|\Psi(\vec{r},t)\rangle, \qquad (1)$$

governs the evolution of the entire state vector $|\Psi(\vec{r},t)\rangle$, any constant (in the configuration space) phase factor of unit modulus (while, otherwise, function of time) is considered, in principle, as unobservable [1]. From the other side, as the interference phenomena takes place within the QM phenomenology, this invokes the quantum superposition principle to hold in the QM. The agreement between the standard interpretation and the superposition principle requires that changing the material phase (MP) of some of the superimposed states will not lead to observable physical consequences. In view of some analytic results and the critical inspection of some matter-wave experiments this, however, appears to be not the case [2]. As has been found, the phase of the state vector depends causally on the initial conditions and the relevant physical processes [2]. What is more important, some experimental results with material wave packets within atoms [3] and molecules [4, 5], and large number of matter-wave experiments [6], while not regarded in the sense discussed here, also confirm such a point of view. Even a *constant phase shift* leads to observable physical results [3, 4]. In this work, we consider the behavior of the MP of the state vector of a quantum system (QS) involved in a real physical process. The problem is treated within the semiclassical dressed states (DSs), adiabatic states, picture. The existence of general relationship between the quantum amplitude and phase phenomena is discussed based on the "hydrodynamic" formulation of QM [7].

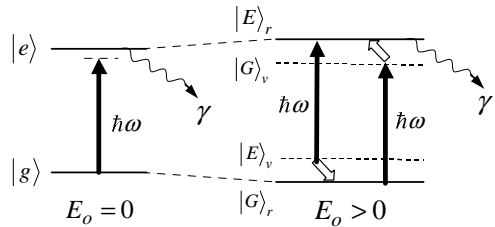

Fig.1. Bare ($E_0=0$) and dressed ($E_0>0$) states of the quantum system. The bold arrows show the optical pumping, the empty arrows show the nonadiabatic processes, and the wavy arrows show the dumping.

Consider the basic model of two-level QS (also regarded as a single quantum bit-qubit), Fig.1, subject to a near-resonant interaction with an electromagnetic field (EMF) and decoherence due to dumping, $\gamma = \gamma' - i\gamma''$. To involve variety of phase contributions, we allow nonadiabatic amplitude $E_o(t)$ and phase $\Phi(t) = \omega t + \varphi(t)$ variations of the EMF, which, however, become constrained by a generalized adiabatic condition of infinite order [2, 8]

$$\left|\partial_t^n(\partial_t\varphi - i\Omega^{-1}\partial_t\Omega)\right| \ll \left|\Delta\omega - i\gamma/2\right|^{n+1-k}|\Omega|^k, \qquad (2)$$

where $(\Delta\omega - i\gamma/2)$ is complex frequency detuning, $\Omega$ is Rabi-frequency, and $n = 0, 1, 2,...; \ k = 0, 1, 2,..., n+1$.

Switching the EMF on creates a superposition state $|\Psi(\vec{r},t)\rangle = g(t)|g(\vec{r})\rangle\exp[-i\Phi_g(t)] + e(t)|e(\vec{r})\rangle\exp[-i\Phi_e(t)]$, where $g(t)$ and $e(t)$ are time-dependent amplitudes of the ground $|g\rangle$ and excited $|e\rangle$ bare states, $\Phi_g(t) = \varphi_g + \int_0^t \omega_g(t')dt'$, $\Phi_e(t) = \varphi_e + \int_0^t \omega_e(t')dt'$ are the total MPs; $\omega_g$, $\omega_e$ are eigenfrequencies. The initial (constant) phases $\varphi_g$ and $\varphi_e$ are considered here at an equivalent ground with the time-dependent parts of the MP due to formal and phenomenological reasons discussed in details in [2]. To derive the DSs we solve Eq. (1) with Hamiltonian $\hat{H} = \hbar\omega_g|g\rangle\langle g| + \hbar\omega_e|e\rangle\langle e| - \mu E(|g\rangle\langle e| + h.c.) - i\hbar(\gamma/2)|e\rangle\langle e|$, which leads to the following solution for the real ("r") and virtual ("v") components of the ground $|G\rangle$ and excited $|E\rangle$ DSs [2]:

$$\begin{aligned}|G\rangle_r &= |g\rangle\exp(-i\Phi_{G,r})\\ |G\rangle_v &= |e\rangle\exp(-i\Phi_{G,v})\\ |E\rangle_r &= |e\rangle\exp(-i\Phi_{E,r})\\ |E\rangle_v &= |g\rangle\exp(-i\Phi_{E,v})\end{aligned} \quad (3)$$

The MP at ground (left) and excited (right) state initial conditions are

$$\begin{aligned}\Phi_{G,r} &= \varphi_g + \int_0^t \omega_G dt' & \Phi_{E,r} &= \varphi_e + \int_0^t \tilde{\omega}'_E dt'\\ \Phi_{G,v} &= (\varphi_g + \varphi) + \int_0^t (\omega_G + \omega)dt' & \Phi_{E,v} &= (\varphi_e - \varphi) + \int_0^t (\tilde{\omega}'_E - \omega)dt'\\ \Phi_{E,r} &= (\varphi_g + \varphi) + \int_0^t \tilde{\omega}'_E dt' & \Phi_{G,r} &= (\varphi_e - \varphi) + \int_0^t \omega_G dt'\\ \Phi_{E,v} &= \varphi_g + \int_0^t (\tilde{\omega}'_E - \omega)dt' & \Phi_{G,v} &= \varphi_e + \int_0^t (\omega_G + \omega)dt'\end{aligned} \quad (4)$$

where $\tilde{\omega}'_E = \omega_E - \partial_t\varphi - \gamma''/2 - i(\gamma'/2 - \Omega^{-1}\partial_t\Omega)$ is the effective frequency of the real excited state; $\omega_G = \omega_g + \Lambda'_-$ and $\omega_E = \omega_e - \Lambda'_-$ are the Stark-shifted frequencies of the real ground and excited states, respectively, $\Lambda_\pm = 1/2(\Delta\tilde{\omega} \pm \tilde{\Omega}')$, $\tilde{\Lambda}'_\pm = \Lambda_\pm - (i/2)\tilde{\Omega}'^{-1}\partial_t\tilde{\Omega}'$, and $\tilde{\Omega}' = [\Delta\tilde{\omega}'^2 + \Omega^2 - i2\partial_t\Delta\tilde{\omega}']^{1/2}$ has meaning of instantaneous off-resonance Rabi frequency.

A number of conclusions can be done from Eqs. (3) and (4), which represent closed form expressions for the internal phase dynamic of single qubit. At ground state initial conditions, the ground state constant MP $\varphi_g$ appears in all DS components, whereas the excited state MP $\varphi_e$ totally disappears. In addition, the phase of the DSs behaves as an *additive dynamical quantity* that causally follows the process of its formation. Really, starting from $|G\rangle_r$ (ground state initial conditions, or $|G\rangle_r \to |G\rangle_v \Rightarrow |E\rangle_r \to |E\rangle_v$ sequence), the phase of the virtual component $|G\rangle_v$ results from the phase of the real component $|G\rangle_r$, adding the optical phase. At the same time, the state $|G\rangle_v$ results physically from $|G\rangle_r$ by temporal association of a field photon. Due to the nonadiabatic coupling between $|G\rangle$ and $|E\rangle$, the phase of $|G\rangle_v$ transfers to $|E\rangle_r$ acquiring nonadiabatic contributions from both, the field ($\partial_t\varphi$, $\Omega^{-1}\partial_t\Omega$) and dumping ($\gamma''/2$, $\gamma'/2$) factors. The variation of the field phase $\partial_t\varphi$ affects the instantaneous energy of the level, whereas the variation of the field amplitude $\Omega^{-1}\partial_t\Omega$ affects the amplitude of the state, populating $|E\rangle_r$ from $|G\rangle_v$. This is accompanied by irreversible absorption of one photon from the field. Finally, the phase of the virtual component $|E\rangle_v$ results from that one of $|E\rangle_r$ subtracting the optical phase. Physically this corresponds to emission of a virtual photon from $|E\rangle_r$ to the field, thus forming $|E\rangle_v$. At excited state initial conditions ($|E\rangle_r \to |E\rangle_v \Rightarrow |G\rangle_r \to |G\rangle_v$ sequence), a similar phase behavior can be found but this time it is the excited state MP $\varphi_e$ appearing in all DSs components, whereas $\varphi_g$ totally disappearing [9]. Consequently, the phase of the DSs (constant and time-dependent) is *causally* related with the initial conditions and the relevant physical processes.

The physically sensible behavior of the MP, and the experimentally observed manifestation of the MP [3-6], show that *the MP is causally involved in the dynamic of the QS*.

One may argue that such MP causality becomes unobservable because, according to the standard interpretation of QM, the entire state vector is unobservable. The inconsistency of such objection becomes clear because of some general arguments. First of all, unobservability of the entire state vector does not necessarily mean unobservability of the MP, or the amplitude, alone. Next, the polar representation of the wave function, $\Psi = R(\vec{r},t)\exp[iS(\vec{r},t)/\hbar]$, leads to the coupled differential equations for the amplitude $R$ and the Hamilton-Jacobi function $S$ [7]

$$\partial S/\partial t + (\nabla S)^2/2m + V(\vec{r},t) + U(\vec{r},t) = 0 \qquad (5)$$

$$\partial (R)^2/\partial t + \nabla.(R^2\vec{v}) = 0 \qquad , \qquad (6)$$

where $\vec{v} = \nabla S/m$ is the velocity of the particle, $V(\vec{r},t)$ is the usual interaction potential, $U(\vec{r},t) = -(\hbar^2/2m)(\Delta R/R)$ is the quantum "potential", which, within the Bohm's interpretation of QM, is considered as a real interaction due to the real physical field $\Psi$ [7]. The first equation represents the quantum Hamilton-Jacobi equation, while the second one becomes the continuity equation and reflects the conservation of the quantum probability density $\rho = R^2$. Because the coupled equations (5) and (6) directly follow from the Schröedinger equation (1), the amplitude $R$ and phase, which actually is $\Phi = -S/\hbar$, are not independent but codetermine each other. Thus, it is hardly to believe that if $R^2$ becomes related to some element of the physical reality, $S$ ($\Phi$) can be irrelevant to the physical reality. Such $R - S$ ($R - \Phi$) relationship exists *independently on the particular interpretation of QM*. Furthermore, whereas the phase $\Phi = -S/\hbar$ is governed by the real *dynamics equation* (5), Eq. (6), which governs the square of the state vector amplitude $R^2$, plays only an auxiliary role related with the probability density definition of the quantum ensemble ($\rho = R^2$) and its conservation. Consequently, the phase $\Phi$ is even more closely related to the dynamic of QS, than the amplitude $R$, that has only an ensemble averaged probability meaning. In that concept, the relation of the phase of the state vector with the quantum dynamic (particle momentum) is given by $\vec{p} = -\hbar\nabla\Phi$.

The ultimate reason to accept the MP causality is the existence of, although sounds strange, a large number of experimental evidences. Thus, gain or quench the population of given quantum state [3-5], or change the fringe pattern [6] in the mater wave interferometers, changing the phase of the interfering material wave packets, are well-established experimental facts. Based on the unity of the physical phenomena, the indisputable influence of the optical phase on the optical wave phenomena can be also considered as evidence on an equivalent ground.

The acceptance of the MP causality does not necessarily require acceptance of the *ontological* interpretation, considering $\Psi$ as an objectively real field [7]. We consider that, while the entire wave function $\Psi$ has not physical meaning, its elements, amplitude $R$ and phase $\Phi$ ($S$), have observable appearances in the physical processes. Such understanding preserves the *epistemological* meaning for the entire state vector.

In summary, the standard interpretation of the quantum mechanics does not comprise the whole quantum mechanical phenomenology. The material phase of a quantum system, which appears to be *a missed parameter* within the standard interpretation, is causally related with the dynamic of the quantum system. The material phase has observable physically consequences and is able to carry physical information.